\documentclass[aps,twocolumn,pra,amsmath,amssymb,floatfix]{revtex4}

\def\ket#1{|#1\rangle}
\def\bra#1{\langle#1|}
\def\av#1{\langle#1\rangle}

\usepackage{graphicx}
\usepackage{natbib}
\usepackage{multirow}

\begin{document}

\title{Vibration-assisted resonance in photosynthetic excitation energy transfer}

\author{E. K. Irish}
\email{eki2@st-andrews.ac.uk}
\affiliation{SUPA, Institute of Photonics and Quantum Sciences, Heriot-Watt University, Edinburgh, EH14 4AS, United Kingdom}
\affiliation{SUPA, Department of Physics and Astronomy, University of St Andrews, KY16 9SS, United Kingdom}

\author{R. G\'{o}mez-Bombarelli}
\affiliation{SUPA, Institute of Photonics and Quantum Sciences, Heriot-Watt University, Edinburgh, EH14 4AS, United Kingdom}

\author{B. W. Lovett}
\email{bwl4@st-andrews.ac.uk}
\affiliation{SUPA, Institute of Photonics and Quantum Sciences, Heriot-Watt University, Edinburgh, EH14 4AS, United Kingdom}
\affiliation{SUPA, Department of Physics and Astronomy, University of St Andrews, KY16 9SS, United Kingdom}
\affiliation{Department of Materials, University of Oxford, Oxford, OX1 3PH, United Kingdom}
\date{\today}

\begin{abstract}
Understanding how the effectiveness of natural photosynthetic energy harvesting systems arises from the interplay between quantum coherence and environmental noise represents a significant challenge for quantum theory. Recently it has begun to be appreciated that discrete molecular vibrational modes may play an important role in the dynamics of such systems. As an alternative to computationally demanding numerical approaches, we present a microscopic mechanism by which intramolecular vibrations contribute to the efficiency and directionality of energy transfer. Excited vibrational states create resonant pathways through the system, supporting fast and efficient energy transport. Vibrational damping together with the natural downhill arrangement of molecular energy levels gives intrinsic directionality to the energy flow. Analytical and numerical results demonstrate a significant enhancement of the efficiency and directionality of energy transport that can be directly related to the existence of resonances between vibrational and excitonic levels.
\end{abstract}

\maketitle

\section{Introduction}

Photosynthetic organisms have evolved a system of light-harvesting antenna complexes that absorb energy from sunlight and funnel it into a reaction centre where the captured light energy is converted into stored chemical energy~\cite{Blankenship:molecularmechanisms}. For this extended system to function effectively, energy transfer through the antenna to the reaction centre must be efficient, in the sense that energy is transferred from molecule to molecule with high probability. The flow of energy also needs to be preferentially directed toward the reaction centre. Understanding how such efficient and directional energy transport arises in natural photosynthetic systems is currently a major area of research.

Two theoretical approaches to studying energy transport are usually distinguished. When the coupling between molecules in a network is weak compared to their interaction with the environment, energy transfer is incoherent and can be described as an effectively classical `hopping' process with rates determined by the Fermi Golden Rule. The opposite limit is the coherent or exciton approach, wherein the molecules are strongly coupled and interact only weakly with their environment. A major theoretical challenge is presented by the fact that many photosynthetic systems, such as the much-studied Fenna-Matthews-Olson complex, fall into an intermediate regime.

In the incoherent F{\"o}rster resonance energy transfer (FRET) model, the conditions for efficient and directional energy transfer are well understood~\cite{Andrews2007}. The F{\"o}rster resonance energy transfer (FRET) mechanism can facilitate efficient energy transfer only between resonant energy levels. In photosynthetic pigment-protein complexes (PPCs), however, chromophores are typically arranged so that their energy levels form a downhill gradient or `energy funnel'~\cite{Blankenship:molecularmechanisms}. Such disordered systems can still support efficient energy transfer thanks to the existence of nuclear vibrational sidebands. Resonant transfer occurs between a donor molecule and a vibrationally excited state of a lower-energy acceptor molecule. The vibrational excitations decay on a relatively fast timescale, creating a difference between the excitation and fluorescence frequencies of the molecule known as the Stokes shift. The Stokes shift favours downhill energy transfer, producing directionality in the FRET mechanism.  

Although nuclear vibrational modes are central to producing efficient one-way energy transport in the incoherent FRET theory, coherent models of energy transfer are usually concerned with the electronic degrees of freedom alone. Vibrations, whether they originate from intramolecular nuclear motion or motions of the protein and solvent environment, are reduced to a collective environment or `bath' which can then be treated at various levels of approximation. In the simplest case, often termed dephasing-assisted transport (DAT) \cite{Plenio2008} or environment-assisted quantum transport (ENAQT) \cite{Rebentrost2009}, the primary effect of the bath is to dephase electronic coherences between different molecules. Studies of this model have produced the important and seemingly counterintuitive insight that environmental noise can improve energy transport in disordered quantum networks~\cite{Leegwater1996,Plenio2008, Rebentrost2009, Caruso2009, Chin2010, Chin2012}. 

As in the incoherent case, efficient energy transfer in coherent models requires resonance between molecules. Consider the simple case of two sites. Fully coherent Hamiltonian dynamics can only produce full population transfer from one site to the other if the two sites have identical energies. Once the energy difference exceeds the coupling, the amplitude of population oscillations is significantly reduced and the excitation becomes primarily localised on a single site. The localisation effects of quantum coherent dynamics in disordered systems can be overcome to some extent by the addition of dephasing. The environmentally induced energy fluctuations responsible for dephasing can momentarily bring the energies of two adjacent sites into resonance, allowing efficient transfer between them. However, since the energy gaps between adjacent chromophores can vary substantially, the individual dephasing rates required to optimise transport without completely destroying coherence must also vary substantially. Studies on the Fenna-Matthews-Olson complex (FMO) found optimised dephasing rates that differ in some cases by two orders of magnitude on adjacent sites~\cite{Plenio2008, Caruso2009}. In a highly structured PPC such as FMO, where the average chromophore separation is on the order of 1.2~nm \cite{Milder2010}, it seems unlikely that adjacent chromophores experience such drastically different noise levels. 

Producing one-way energy transport in coherent models is more difficult. Fully coherent Hamiltonian evolution is inherently reversible. Open quantum systems techniques overcome this problem by taking the quantum system under study to be coupled to a large environment, leading to effectively irreversible evolution. Such evolution may or may not appear as a directional, coherent energy transfer process. Pure dephasing, as in DAT/ENAQT models, results in a diffusive process whose limiting distribution is equal population on each site in the system~\cite{Rebentrost2009}. In order to achieve one-way energy transport, these models rely on the addition of a trap site to which energy is transferred irreversibly at a constant rate. This mechanism is designed to model exciton transfer from peripheral complexes to the reaction centre~\cite{Leegwater1996}. In the case of the much-studied FMO complex, there is in fact very little experimental data on which to base the trapping model~\citep[see][and references therein]{Amesz2002}. Furthermore, the key experiments on energy transfer dynamics in FMO studied purified complexes that do not contain reaction centres, and thus any observed directionality must originate from a different mechanism. 

The Stokes shift together with the downhill arrangement of energy levels provides directionality in the incoherent FRET model. However, the level of approximation employed in deriving pure dephasing models eliminates the Stokes shift. The Lindblad master equation with pure dephasing is equivalent to the stochastic Haken-Strobl model~\cite{Rebentrost2009}. The latter corresponds to the fast modulation or high temperature limit, in which the bath a has zero correlation time~\cite{Rips1993}, and the Stokes shift vanishes in this limit~\cite{Mukamel:nonlinear}. Leegwater~\cite{Leegwater1996} has further shown that coherent dynamics with pure dephasing is identical to the incoherent hopping predicted by FRET when the interaction with the environment becomes much larger than the coherent coupling between individual molecules. In the high-temperature limit required for the bath correlation time to vanish, the FRET rates for forward and backward energy transfer become equal and there is no longer a preferred transfer direction.

Recent work, both experimental and theoretical, has shown that vibrations in PPCs do not behave like simple thermal bath with a smooth spectral density. Vibrational spectroscopy has revealed rich intramolecular vibrational structures in FMO~\cite{Wendling2000,Ratsep2007} and other photosynthetic complexes~\cite{Doust2004, Novoderezhkin2010}. Molecular dynamics simulations of FMO have likewise shown that the spectral density contains a number of distinct peaks that can be attributed to intramolecular nuclear vibrational modes~\cite{Olbrich2011c,Shim2012}. Some recent numerical studies have suggested that these modes may play a substantial role in coherent energy transfer~\cite{Chin2010, Ritschel2011, delRey2013, Chin2013}, particularly as a number of vibrational frequencies lie within the range of exciton energy splittings and resonance effects may therefore be important~\cite{Lim2013, Kolli2012, Chenu2012, Chin2012, delRey2013}. However, the microscopic mechanisms by which vibrations can contribute to energy transfer are not well understood and the specific role of resonance has yet to be conclusively demonstrated. 

The theory introduced here provides an alternative approach to satisfying the conditions of resonance and directionality, by including nuclear vibrational modes explicitly in the coherent part of the system under study. As in FRET, resonant transfer occurs between a donor molecule and a vibrationally excited state of the acceptor molecule. Subsequent decay of the vibrational excitation effectively produces a Stokes shift, creating a preferential direction of transfer. However, in our model both the electronic states and the relevant vibrational states are treated coherently within an exciton approach. The environmental interaction is still described by a Lindblad master equation, as in DAT/ENAQT-type models, retaining much of the simplicity of pure dephasing models while incorporating critical features of nuclear vibrational coupling.

\section{Results}

\subsection{A coherent model for vibration-assisted resonance}

The starting point for our theory is the standard electronic Hamiltonian $H_{\text{el}}$ used in exciton theory~\cite{Renger2001}. In the site basis, the electronic Hamiltonian is given by
\begin{equation}
H_{\text{el}} = \sum_{i=1}^N \epsilon_i \ket{i}\bra{i} + \sum_{i \neq j} J_{ij} \ket{i}\bra{j}
\label{eq:Hel}
\end{equation}
where $N$ is the number of chromophores or sites, $\ket{i}$ denotes the electronic excited state of molecule $i$, $\epsilon_i$ is the electronic excitation energy of $\ket{i}$ relative to its ground state $\ket{g_i}$, and $J_{ij}$ is the excitonic coupling between sites $i$ and $j$. 

To the electronic Hamiltonian we add two additional terms, $H_{\text{vib}}$ and $H_{\text{el-vib}}$, to describe the vibrational modes and their coupling to the electronic states, respectively. The relevant vibrational modes are intramolecular vibrations excited by the electronic transition from ground to excited state~\cite{Turro:Modern}. Nonlinear polyatomic molecules with \textit{N} atoms possess 3\textit{N}-6 nuclear degrees of freedom excluding rotations (in the Eckart conditions) and translations. These motions - molecular vibrations - are usually described within the harmonic approximation as small displacements on a parabolic potential energy surface around an equilibrium position. When a chromophore absorbs an incoming photon, an electron is promoted from an occupied low-energy molecular orbital to an unoccupied one with higher energy. The excitation is much faster than nuclear motions and thus, within the Franck-Condon approximation, it is assumed that nuclei remain static during the optical excitation and afterwards relax to the excited-state equilibrium geometry. To lowest order, the excited state normal modes can be approximated by the same parabolic potential energy surfaces as in the ground state but with a displaced equilibrium position. A similar process occurs during fluorescence emission: the nuclei remain static during the transition - in this case at the equilibrium geometry of the excited electronic state - and after the change in electronic state, vibrational modes mediate the relaxation to the ground state equilibrium geometry. Experimentally, the energy difference between absorption and emission peaks in the spectra is termed the Stokes shift.

The vibrational Hamiltonian consists of $n(i)$ vibrational modes of each of the $i$ molecules:
\begin{equation}
H_{\text{vib}} = \sum_{i=1}^N \sum_{k=1}^{n(i)} \omega_{ik} a^{\dag}_{ik}a_{ik}
\label{eq:Hvib}
\end{equation}
where $a^{\dag}_{ik}$ ($a_{ik}$) is the raising (lowering) operator for the mode with frequency $\omega_{ik}$. The zero-point energy has been omitted and $\hbar$ has been set to $1$. Each vibrational mode is linearly coupled with strength $\lambda_{ik}$ to the excited state of the chromophore on which the vibration is localised, giving
\begin{equation}
H_{\text{el-vib}} = \sum_{i=1}^N \sum_{k=1}^{n(i)} \lambda_{ik} (a^{\dag}_{ik} + a_{ik}) \ket{i}\bra{i} .
\label{eq:Helvib}
\end{equation}
We have assumed that the Franck-Condon approximation is valid and that the ground- and excited-state modes differ only by a displacement in equilibrium position. Written in the ground-state basis, the displacement of the corresponding excited-state vibrational mode appears as a linear coupling term.

Interactions with the environment, comprised of protein vibrational modes and solvent fluctuations, are incorporated by means of a Markovian master equation~\cite{Breuer:openquantumsystems}:
\begin{equation}\label{eq:mastereq}
\frac{d \rho}{d t} = -i[H, \rho(t)] - \mathcal{L}_{\text{deph}}(\rho(t)) - \mathcal{L}_{\text{vib}}(\rho(t)),
\end{equation}
where $H = H_{\text{el}} + H_{\text{vib}} + H_{\text{el-vib}}$ and $\rho(t)$ is the density matrix of the full system of electronic states and nuclear vibrational modes. Dephasing of electronic coherences is described by the Lindblad superoperator
\begin{equation}
\mathcal{L}_{\text{deph}}(\rho(t)) = \sum_{i=1}^N \gamma_i^{\text{deph}} (\ket{g_i}\bra{g_i} - \ket{i}\bra{i}) ,
\end{equation}
where $\ket{g_i}$ denotes the ground state of site $i$ and $\gamma_i^{\text{deph}}$ is the electronic dephasing rate of site $i$. Damping of the nuclear vibrational modes is needed to incorporate a Stokes-shift-like effect into the exciton model and is given by
\begin{equation}
\begin{split}
\mathcal{L}_{\text{vib}}(\rho(t)) = \sum_{i=1}^N \sum_{k=1}^{n(i)} \gamma_{ik}^{\text{vib}} [&(\nu_{ik} + 1) (\{ a^{\dag} a, \rho \} - 2 a \rho a^{\dag}) \\
&  + \nu_{ik}(\{ a a^{\dag}, \rho \} - 2 a^{\dag} \rho a )] .
\end{split}
\end{equation}
Here $\{A, B\} = AB + BA$ denotes the anticommutator, $\gamma_{ik}^{\text{vib}}$ is the damping rate of mode $ik$, and $\nu_{ik}$ is the number of thermal excitations in mode $ik$ in the steady state. We have omitted decay of the electronic excitations since we are interested in dynamics on time scales much shorter than the electronic decay time, but this effect is easily incorporated within the master equation formalism.

In realistic biological systems, the addition of intramolecular vibrational modes to the system Hamiltonian rapidly expands the system size beyond the limits of computational feasibility. Even a system as small as FMO has around thirty strong vibrational modes~\cite{Wendling2000,Ratsep2007}, making a full system treatment intractable. However, we argue here that only the vibrational levels that create resonant pathways through the network contribute substantially to the dynamics of energy transfer. With this restriction the system size can be substantially reduced while retaining the essential physics. 

\subsection{Analysis of a simple case}

\begin{figure}[t]
\includegraphics[scale=.47]{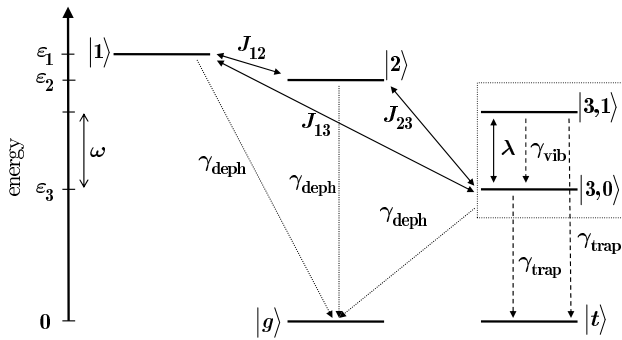}
\caption{Diagram of the three-site model, with site 3 coupled to a vibrational mode that is restricted to either 0 or 1 excitation(s). Solid arrows indicate coherent couplings, dashed arrows indicate irreversible decay pathways, and dotted arrows indicate dephasing pathways.}
\label{fig:model}
\end{figure}

To illustrate the vibration-assisted resonance mechanism we construct a minimal model, based on sites 1-3 of the FMO complex, in which the relevant physical effects can be clearly seen. To a good approximation, these three sites comprise one of two energy transfer branches identified in FMO~\cite{Adolphs2006, Brixner2005}. The model, shown in Fig.~\ref{fig:model}, consists of three sites, whose electronically excited states are denoted by $\ket{i}$, $i=1,2,3$; the overall ground state of the system is denoted by $\ket{g}$. Sites 1 and 2 are near resonance and strongly coupled, while site 3 has a much smaller excitation energy and is more weakly coupled to the other sites. For simplicity we include only a single vibrational mode of frequency $\omega$, linearly coupled with strength $\lambda$ to site 3. Furthermore, we restrict the mode to its ground and first excited states; these states are denoted, respectively, by $\ket{3,0}$ and $\ket{3,1}$. Equations~\eqref{eq:Hel}-\eqref{eq:Helvib} then reduce to
\begin{equation}\label{eq:Hamiltonian}
\begin{split}
H &= \epsilon_1 \ket{1}\bra{1} + \epsilon_2 \ket{2}\bra{2} + \epsilon_3 \ket{3,0}\bra{3,0} \\
&  \quad + (\epsilon_3 + \omega) \ket{3,1}\bra{3,1} \\
& \quad + J_{12}(\ket{1}\bra{2} + \ket{2}\bra{1}) + J_{23}(\ket{2}\bra{3,0} + \ket{3,0}\bra{2}) \\
& \quad + J_{13}(\ket{1}\bra{3,0} + \ket{3,0}\bra{1}) \\
& \quad + \lambda(\ket{3,0}\bra{3,1} + \ket{3,1}\bra{3,0}) .
\end{split}
\end{equation}
The dephasing rate $\gamma_{\text{deph}}$ is assumed to be the same for each site, and the vibrational mode decays from its excited state $\ket{3,1}$ to its ground state $\ket{3,0}$ with rate $\gamma_{\text{vib}}$. In order to compare our results directly with those from DAT/ENAQT models we add a trapping state $\ket{t}$ that absorbs energy from both vibronic levels on site 3 at rate $\gamma_\text{trap}$. We also consider the case without the trap and show that it is not, in fact, necessary to produce directional energy transfer. For simplicity we neglect losses from exciton recombination or other non-radiative decay processes which occur on nanosecond timescales~\cite{Milder2010}, much slower than the $0.5-5~\text{ps}$ timescales we are interested in.

To ensure that our model is both physically motivated and realistic, parameter values used in the discussion of the model and numerical simulations are drawn from experimental results wherever possible. Exact values for the site energies and excitonic couplings vary depending on the organism being studied and the details of experimental spectra and theoretical fitting methods~\cite{Milder2010}. However, certain trends are clear. The relations $\epsilon_1 \sim \epsilon_2 > \epsilon_3$, $J_{12} > J_{23} > J_{13}$, $J_{12} \gg |\epsilon_1 - \epsilon_2|$ and $\epsilon_2 - \epsilon_3 \gg J_{23}$ are useful in obtaining physical insight into the model. For numerical calculations we use site energies and excitonic couplings from \cite{Adolphs2006}, as shown in Table~\ref{tab:energycoupling}. 

\begin{table}[ht]
\begin{tabular}{cr|cr}
\hline
\multicolumn{2}{r|}{Site energies} & \multicolumn{2}{c}{Couplings} \\
\hline
$\epsilon_1$ & 12475 & $ J_{12}$ & -98.2 \\
$\epsilon_2$ & 12460 & $ J_{13}$ & 5.4 \\
$\epsilon_3$ & 12225 & $ J_{23}$ & 30.5 \\
\hline
\end{tabular}
\caption{Parameter values in $\text{cm}^{-1}$ for \textit{Prosthecochloris aestuarii}, taken from \cite{Adolphs2006}.}
\label{tab:energycoupling}
\end{table}

The magnitude of the vibrational coupling constant $\lambda$ requires careful consideration. This term originates from the shift of the nuclear normal-mode potential energy surface upon electronic excitation of a pigment molecule~\cite{Renger2001,Jean1992}, from which $\lambda = \omega \sqrt{S_{\omega}}$ where $S_{\omega}$ is the Huang-Rhys factor of the vibrational mode with frequency $\omega$. Adolphs and Renger~\cite{Adolphs2006} considered a vibrational spectral density consisting of a broad continuous background with a single high-frequency mode at $\omega_H = 180~\text{cm}^{-1}$. Comparing this model with the experimental vibrational spectra of Wendling ~\textit{et al.}~\cite{Wendling2000}, they estimated the Huang-Rhys factor $S_H = 0.22$. Subsequent work has used the same or similar values~\cite{Chin2010,Chin2012}. However, the Huang-Rhys factors for individual modes found in the experiments of Wendling ~\textit{et al.} are on the order of $S_\omega \lesssim 0.01$, and in fact the sum of the Huang-Rhys factors for the thirty measured vibrational modes is on the order of $S_\text{tot} \sim 0.25$~\cite{Wendling2000}. As the physics presented here is based on a resonance mechanism and therefore relies on close frequency matching, we choose the Huang-Rhys factor $S_\omega \sim 0.01$ appropriate to an individual mode rather than the value $S \sim 0.2$ that corresponds to an effective coupling to a large number of modes. For the frequencies of interest in our model this yields $\lambda \sim 15~\text{cm}^{-1}$, so that $\lambda < J_{23} < J_{12}$.

\begin{figure}[t]
\includegraphics[scale=.5]{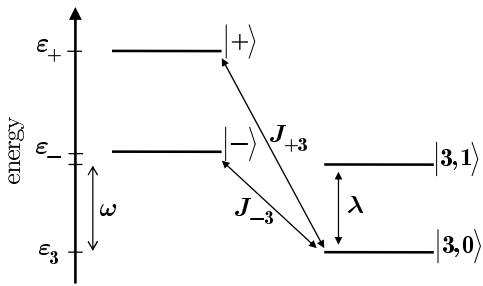}
\caption{Simplified model illustrating the resonance mechanism within a perturbative framework. The delocalised excitonic states $\ket{\pm}$ diagonalise the $\{\ket{1},\ket{2}\}$  subspace. The transformed couplings $J_{\pm 3}$ are of similar magnitude to $\lambda$, which is much smaller than $\epsilon_{\pm} - \epsilon_3$. When $\omega \simeq \epsilon_{\pm} - \epsilon_3$, a resonance occurs and the exciton state $\ket{\pm}$ becomes strongly mixed with the vibrationally excited state $\ket{3,1}$ even though the states are only coupled to second order (via the intermediate state $\ket{3,0}$) and both $J_{\pm 3}$ and $\lambda$ are small. }
\label{fig:pertmodel}
\end{figure}

The first step in our analysis is to identify the resonances in the system and their effect on the system in the absence of any environmental interactions. From the relations among the various energies and coupling strengths discussed above, we can extract a simplified picture that provides good intuition for the physics of the model. States $\ket{1}$ and $\ket{2}$ are near resonance and strongly coupled, so they must be treated in the delocalized `exciton basis'. This produces the set of basis states and couplings illustrated in Fig.~\ref{fig:pertmodel}: the new states $\ket{\pm}$ with energies $\epsilon_{\pm}$ are the excitonic states spanning the $\{\ket{1},\ket{2}\}$ subspace, $J_{\pm 3}$ denotes their couplings to state $\ket{3,0}$, and the coupling $\lambda$ between $\ket{3,0}$ and $\ket{3,1}$ remains unchanged. For our system $|J_{\pm 3}| \gtrsim \lambda$ and all three couplings are small enough to be treated as perturbative parameters. As long as $\omega$ is chosen so that $\ket{3,1}$ does not come into resonance with $\ket{\pm}$, the eigenstates will not vary significantly from the basis states. However, when $\omega \simeq \epsilon_{\pm} - \epsilon_3$ a resonance is created and $\ket{3,1}$ can become strongly mixed with $\ket{\pm}$. 

The effect of the resonance may be analysed within this picture using degenerate perturbation theory. The Hamiltonian is divided into a `bare' term $H_0$ and a perturbation term $H^{\prime}$:
\begin{align}\label{eq:pertham}
H &= H_0 + H^{\prime} , \\
\begin{split}
H_0 &= \epsilon_+ \ket{+}\bra{+} + \epsilon_- \ket{-}\bra{-} + \epsilon_3 \ket{3,0}\bra{3,0} \\
&\quad + (\epsilon_3 + \omega) \ket{3,1}\bra{3,1} ,
\end{split} \\
\begin{split}
H^{\prime} &= J_{+3}(\ket{+}\bra{3,0} + \ket{3,0}\bra{+}) \\
& \quad + J_{-3}(\ket{-}\bra{3,0} + \ket{3,0}\bra{-}) \\
& \quad + \lambda(\ket{3,0}\bra{3,1} + \ket{3,1}\bra{3,0}) .
\end{split}
\end{align}
For notational simplicity we will take the resonant pair of states to be $\ket{-}$ and $\ket{3,1}$, so that $\omega = \epsilon_- - \epsilon_3$. The resulting degeneracy in $H_0$ creates divergences in the series expansions of the eigenstates of $H$. Therefore the first step in applying perturbation theory must be to identify the appropriate superpositions of the resonant states that will remove the degeneracy.

Looking at Eq.~\eqref{eq:pertham} it is immediately evident that there is no matrix element of $H^{\prime}$ that directly connects the two resonant states, so the degeneracy cannot be removed by simply diagonalising the degenerate subspace of $H$. It is necessary, then, to work to second order in the expansion of the eigenstates and energies~\cite{Schiff:quantummechanics} to determine an \textit{effective} perturbation Hamiltonian $\tilde{H}$ in the $\{\ket{-}, \ket{3,1}\}$ subspace. For two states $\ket{m}$ and $\ket{l}$ with $E_m = E_l$ the general form of $\tilde{H}$ may be expressed as
\begin{equation}\label{eq:Heff2ndgen}
\tilde{H} = \tilde{H}_{mm} \ket{m}\bra{m} + \tilde{H}_{ll} \ket{l}\bra{l} + (\tilde{H}_{ml} \ket{m}\bra{l} + \text{H.c.}) ,
\end{equation} 
where H.c. denotes Hermitian conjugate and the matrix elements are given by
\begin{align}\label{eq:Heff2ndelements}
\tilde{H}_{mm} &= \sum_{n \neq l,m} \frac{\lvert \av{n | H^{\prime} | m} \rvert ^2}{E_m - E_n} ,\\
\tilde{H}_{ll} &= \sum_{n \neq l,m} \frac{\lvert \av{n | H^{\prime} | l} \rvert ^2}{E_m - E_n} ,\\
\tilde{H}_{ml} &= \sum_{n \neq l,m} \frac{\av{m | H^{\prime} | n} \av{n | H^{\prime} | l}}{E_m - E_n} .
\end{align}
The sum is taken over the remaining nondegenerate states of the bare Hamiltonian. By diagonalising $\tilde{H}$ and taking its eigenstates as the new basis states for the perturbation calculation, the degeneracy that would otherwise cause the expansions to diverge may be removed. 

In the present case each of the degenerate states couples only to $\ket{3,0}$ so that the sums reduce to a single term each and we obtain
\begin{equation}\label{eq:Heff2nd}
\begin{split}
\tilde{H} &= \frac{\lvert J_{-3} \rvert ^2}{\omega} \ket{-}\bra{-} + \frac{\lvert \lambda \rvert ^2}{\omega} \ket{3,1}\bra{3,1} \\
& \quad + \frac{J_{-3} \lambda}{\omega}(\ket{-}\bra{3,1} + \ket{3,1}\bra{-}) ,
\end{split}
\end{equation}
where $J_{-3} = \av{3,0 | H^{\prime} | -} = c_1^- J_{13} + c_2^- J_{23}$ is the coupling between the state $\ket{-} = c_1^- \ket{1} + c_2^- \ket{2}$ and the intermediate state $\ket{3,0}$. 

Inspection of Eq.~\eqref{eq:Heff2nd} shows that, in fact, $\omega = \epsilon_- - \epsilon_3$ is not the exact resonance frequency of the interacting system unless $\lvert J_{-3} \rvert^2 = \lvert \lambda \rvert^2$. The interaction of $\ket{-}$ ($\ket{3,1}$) with $\ket{3,0}$ shifts its energy, and if the interaction strengths are different the states will be shifted slightly out of resonance. This effect may be accounted for by adding a further perturbation term 
\begin{equation}
H^{\prime \prime} = \delta\omega \ket{3,1}\bra{3,1} 
\end{equation}
that allows the resonance frequency $\omega$ to be adjusted. This term acts only to second order in the perturbation expansion, so it does not correct the degeneracy to first order. To second order the effective Hamiltonian becomes
\begin{equation}\label{eq:Heff2ndplusdelta}
\begin{split}
\tilde{H} &= \frac{\lvert J_{-3} \rvert ^2}{\omega} \ket{-}\bra{-} + \left(\frac{\lvert \lambda \rvert ^2}{\omega} + \delta\omega \right)\ket{3,1}\bra{3,1} \\
& \quad + \frac{J_{-3} \lambda}{\omega}(\ket{-}\bra{3,1} + \ket{3,1}\bra{-}) .
\end{split}
\end{equation}
The condition for exact resonance is that the diagonal terms of $\tilde{H}$ are equal, which occurs when
\begin{equation}\label{eq:resfreqcorr}
\delta\omega = \frac{\lvert J_{-3} \rvert ^2 - \lvert \lambda \rvert ^2}{\omega} .
\end{equation}
At this point the energy corrections due to interactions within the degenerate subspace are $\tilde{E}_1 = -\tilde{E}_2 = J_{-3} \lambda/\omega$ and the original basis states $\ket{-}$ and $\ket{3,1}$ are maximally mixed. To get an idea of how well our degenerate perturbation theory analysis works for the FMO complex, we can compare it against a numerical solution of the full system. The resonance frequencies may be found by using the simulations to identify the avoided crossings of the eigenvalues of $H$ as a function of $\omega$, which gives $\omega \approx 147.1~\text{cm}^{-1}$ for the lower resonance. Perturbation theory predicts a resonance frequency $\omega + \delta\omega = 147.1~\text{cm}^{-1}$ in excellent agreement with the numerical result. Likewise the energy splitting $2J_{-3} \lambda/\omega = 5.4~\text{cm}^{-1}$, which agrees well with the numerical values of $5.1~\text{cm}^{-1}$.

The same analysis may be applied when $\omega$ is such that $\ket{3,1}$ is on resonance with the upper excitonic state $\ket{+}$. A prediction of $\omega + \delta\omega = 341.2~\text{cm}^{-1}$ is obtained. However, in this case the couplings $J_{13}$ and $J_{23}$ add destructively, giving $J_{+3} = -16.8~\text{cm}^{-1}$, somewhat smaller than $J_{-3} = 25.9~\text{cm}^{-1}$. Together with the larger resonance frequency, this produces an energy splitting of just $0.7~\text{cm}^{-1}$. Consequentially the effect of the resonance with $\ket{+}$ on the dynamics of the system is much smaller than that of $\ket{-}$. This is not inconsistent with the relaxation pathways proposed by Brixner \textit{et al.} \cite{Brixner2005}, in which exciton 7 (roughly equivalent to $\ket{+}$ in our model) decays to exciton 3 ($\ket{-}$ in our model) rather than directly to the lowest level primarily localised on site 3. Given the small energy difference and large coupling between sites 1 and 2, this transition seems unlikely to be vibrationally assisted in the coherent sense discussed here. Alternatively, Chin \textit{et al.}~\cite{Chin2010} suggest that $\ket{+}$ may couple strongly to one of the exciton states on sites 4-7 of FMO, so it may participate in a more complicated energy transfer mechanism that is not captured in our simplified model. Therefore we will focus on the resonance with $\ket{-}$ in the dynamical simulations to follow.

Despite its apparent simplicity, the model we have constructed captures the interplay between delocalised exciton states and local vibrational modes. It is worth emphasising that the coupling of the $\ket{-}$ and $\ket{+}$ excitonic states to the vibrationally excited state $\ket{3,1}$ is weak, as is the coupling between $\ket{3,0}$ and $\ket{3,1}$. Nevertheless, when the resonance condition is satisfied the exciton state becomes strongly mixed with the vibrationally excited state, creating delocalised vibronic states. The analytical framework provides both an intuitive picture for how resonance can contribute to energy transport and a solid mathematical method for accurately calculating resonance frequencies and the consequent splitting of the vibronic eigenstates. Within this model we can now study how vibronic resonances together with electronic coupling and environmental noise can contribute to energy transfer processes.

\subsection{Dynamics of energy transfer}

Without dephasing or trapping it should be immediately clear that adding a resonant vibrational level provides a dramatic improvement in energy transport across the system. In the absence of the vibrational mode, the large energy gap and small coupling between site 3 and the rest of the system means that initial excitation of site 1 produces an oscillatory population on site 3 with a maximum amplitude of about 2.5\%. In our model the effective Hamiltonian given in Eq.~\eqref{eq:Heff2ndplusdelta} suggests that an excitation in $\ket{-}$ will be completely transferred to state $\ket{3,1}$ at time $\tau = (\pi/2)(\hbar \omega / J \lambda) \approx 3~\text{ps}$, which is of the right order to be of biological relevance. Adding decay to the vibrational mode provides directionality of energy transport. On resonance and with $\gamma_{\text{vib}} = 7.5~\text{cm}^{-1}$ (based on the experimental results of Wendling~\textit{et al.} \cite{Wendling2000}), the population of $\ket{3,0}$ reaches 40\% after 5~ps following an initial excitation on site 1. The final population is primarily limited by the amplitude of $\ket{-}$ in the initial state, as $\ket{+}$ couples very weakly to $\ket{3,0}$.

\begin{figure}[htpb]
\includegraphics[scale=.95]{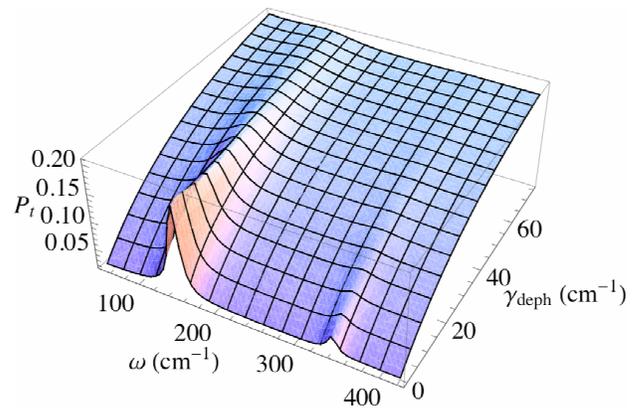}
\caption{Trap state population $P_{\text{t}}$ at 5~ps for an initial state on site 1 as a function of vibrational frequency $\omega$ and dephasing rate $\gamma_{\text{deph}}$ (both measured in $\text{cm}^{-1}$). The trapping rate has been set to $\gamma_{\text{trap}} = 1~\text{cm}^{-1}$. Peaks at $\omega \approx 147~\text{cm}^{-1}$ and $341~\text{cm}^{-1}$ correspond to resonances of $\ket{-}$ and $\ket{+}$ with $\ket{3,1}$.}
\label{fig:results1}
\end{figure}

Biological systems, of course, operate under conditions in which dephasing cannot be neglected. In order to directly compare the vibrationally assisted system with the results of DAT/ENAQT, a trapping process as illustrated in Fig.~\ref{fig:model} was added to the general model. Eq.~\eqref{eq:mastereq} was used to calculate the population $P_{\text{t}}$ of the trap site $\ket{t}$ at time $t = 5~\text{ps}$, following an initial excitation in state $\ket{1}$. The value of $\gamma_{\text{trap}} = 1~\text{cm}^{-1} = 0.03~\text{ps}^{-1}$ was chosen to be similar to that of Refs.~\cite{Plenio2008, Rebentrost2009}. Figure \ref{fig:results1} shows the results as a function of the vibrational frequency $\omega$ and the dephasing rate $\gamma_{\text{deph}}$. At small dephasing, a significant enhancement of the trap population is found for $\omega \sim 147~\text{cm}^{-1}$, corresponding to the resonance between $\ket{-}$ and $\ket{3,1}$. A smaller peak due to the resonance with $\ket{+}$ is visible around $\omega = 341~\text{cm}^{-1}$. As the dephasing increases the effect of the resonance diminishes but remains visible up to the highest values considered. 

Irreversible decay to a trapping site is included to ensure efficiency and directionality within pure dephasing models~\cite{Plenio2008, Rebentrost2009, Chin2010, Chin2012}. Otherwise the system evolves smoothly to a steady-state population given by $1/N$ where $N$ is the number of sites in the network~\cite{Rebentrost2009}. The inclusion of this trapping process is usually justified by the belief that the FMO protein functions as a `molecular wire' that transmits energy from the chlorosome antenna complex to the reaction center~\cite{Olson2004}. Unfortunately there is very little data about the kinetics of energy transfer from FMO to the reaction center; what few experiments exist have shown surprisingly slow and inefficient transfer, in contradiction to expectations from structural evidence~\cite{Amesz2002}. Experiments on other antenna-reaction center complexes have shown that traps spend some fraction of time closed to further energy transfer and also that excitons may have a nontrivial probability of escaping from the trap back into the antenna~\cite{Blankenship:molecularmechanisms, Amesz2002}. In FMO neither the trapping rate nor the nature of the trapping process is known.

\begin{figure*}[htb]
\includegraphics[scale=1.]{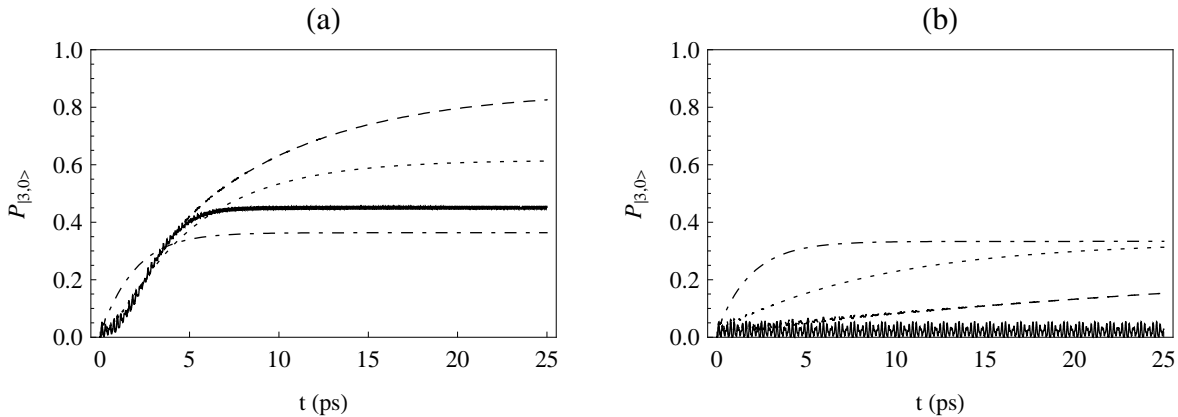}
\caption{Population of $\ket{3,0}$ ($P_{\ket{3,0}})$ in the absence of trapping for an initial excitation on site 1, both with (a) and without (b) vibrational coupling. Dephasing values (in $\text{cm}^{-1}$) are $\gamma_{\text{deph}} =$ 0 (solid), 2 (dashed), 10 (dotted), and 50 (dot-dashed). A longer timescale of 25~ps has been chosen to show the steady-state behaviour of the population. }
\label{fig:notrap}
\end{figure*}

By contrast, the vibration-assisted resonance mechanism produces directional energy transfer without an additional trapping process. The lowest energy site serves as the energy `sink', with the decay of the vibrational level providing the asymmetry between forward and backward transfer. Figure~\ref{fig:notrap} shows the population of $\ket{3,0}$ in the absence of trapping both with (a) and without (b) vibrational coupling. Without the vibrational coupling, setting $\gamma_{\text{deph}} = 0$ demonstrates the poor energy transport properties of purely coherent dynamics in a disordered network. Finite dephasing values all lead to a steady-state population of $1/3$; the larger the dephasing, the faster this state is achieved. In the vibrational case with $\gamma_{\text{deph}} = 0$, the steady-state population is determined by the amplitude of $\ket{-}$ in the initial state $\ket{1}$, here nearly one-half. Adding a small amount of dephasing allows the initial population in $\ket{+}$ to be transferred with high probability. However, increasing the dephasing beyond a certain optimal value (here around $2~\text{cm}^{-1}$) reduces the population transfer. At sufficiently large dephasing values the diffusion limit is recovered, in which the population of $\ket{3,0}$ rises quickly to its steady-state value. The vibration-assisted increase in transfer efficiency appears to be quite sensitive to the precise dephasing rate and approaches its maximum value over a relatively long timescale. On the other hand, the improvement is caused by the inclusion of the decay term on the vibrational mode, a process that has a sound physical basis. It is interesting to note that varying $\gamma_{\text{vib}}$ (not shown) indicates that, on resonance, the energy transfer is optimised at a value very close to the linewidth of $7.5~\text{cm}^{-1}$ obtained in vibrational spectroscopy experiments~\cite{Wendling2000}. 

\section{Discussion}

Drawing on physical intuition from incoherent FRET theory, we have proposed a general model for energy transport dynamics in the coherent regime where internal exciton couplings exceed the interaction with the thermal environment. Incorporating resonant vibrational modes into the system Hamiltonian provides a physically motivated mechanism for simultaneously providing directionality to the energy transport and improving its efficiency over pure-dephasing models. This mechanism does not require the assumption of external energy trapping, individually optimised site dephasing rates, or correlated bath fluctuations. 

Analysis of a simple case based on the FMO complex clearly shows that it is only the resonant vibrational levels that significantly affect the dynamics. Interactions with off-resonant modes renormalise the exciton energies, altering the resonance frequencies of the network but not the character of the eigenstates. This insight suggests a method for applying the general model to specific systems of interest. First, resonances in the excitonic system are identified using perturbation theory as a guide. Only resonant vibrational levels need be included in the system Hamiltonian along with the electronic energies and couplings. Dynamics can then be calculated using a Lindblad master equation that incorporates both electronic dephasing and decay of vibrational excitations. While it is not possible to directly compare the simplified model of FMO treated here with experimental results, a few remarks are in order. 

Comparison of the resonant frequencies obtained in our three-site model with the vibrational frequencies obtained in the experiments of Wendling~\textit{et al.} \cite{Wendling2000} shows that our calculated frequencies actually fall in gaps in the vibrational spectrum. To a certain extent this is because we consider only three sites and a single mode; coupling to the remaining sites and modes provides further shifts to the exciton levels $\ket{\pm}$, which will alter the values of $\omega$. More fundamentally, the resonance frequency depends strongly on the particular values chosen for the site energies and exciton couplings. As yet there is no direct experimental method for measuring the site energies, and values based on theoretical fits to experimental spectra vary considerably~\cite{Milder2010}. In order to estimate the sensitivity of the frequency to changes in the Hamiltonian, we calculated resonance frequencies from a number of published data sets for the site energies and exciton couplings in the FMO complex of \textit{P. aestuarii}. Values for the lower resonance ranged from $94$ to $167~\text{cm}^{-1}$. From this exercise we conclude that the site energies, in particular, of FMO are not yet sufficiently well known to enable accurate calculation of resonance frequencies.

Recent experiments have been designed to address the hypothesis that vibrational and/or vibronic coherence might be responsible for the observed quantum beating in FMO. Hayes~\textit{et al.} studied FMO complexes with modified vibronic structure and found little effect on the frequency or dephasing of the beating in the exciton 1-2 crosspeak \cite{Hayes2011a}. The beat frequency is equal to the energy difference between the two exciton levels. In our model, resonance with the vibrational mode has little effect on the energy difference since the splitting due to the resonance is small, about $5~\text{cm}^{-1}$. Therefore we would not expect to observe much change in the coherence beat frequency when the vibrational mode is shifted off resonance. The primary effect of the resonance is an enhancement of the energy transfer efficiency, which was not addressed in that particular experiment. Another experiment designed to compare the 2D spectra of Bchl~\textit{a} in solution with the spectra of FMO was unable to resolve more than two vibrational modes, which had frequencies well above those relevant for our model \cite{Fransted2012}. Neither of these experiments would seem to rule out a vibration-assisted resonance mechanism.

On the other hand, an experiment examining the inhomogeneous broadening of the exciton 1-3 coherence in FMO found the energy gap between the exciton levels to be quite consistent across the sample~\cite{Fidler2012}. This is not necessarily a general feature of photosynthetic pigment-protein complexes, as the authors also studied the LH2 antenna complex from purple bacteria and found a large degree of inhomogeneous broadening. Their result supports the idea that having a well-defined exciton splitting is important to the functioning of the FMO complex. Such an interpretation would be more consistent with our model, which requires a close match between the exciton splitting and a specific vibrational frequency, than with the dephasing-assisted mechanism that places no restrictions on the exciton splittings. Ultimately, though, determining whether coherent vibration-assisted resonance plays a significant role in biological function will require a better understanding of the kinetics and decoherence properties of pigment-protein complexes.

Finally, we speculate that the vibration-assisted resonance mechanism may be able to shed some light on the current controversies over experiments demonstrating long-lived coherence oscillations in the FMO complex. The model suggests that the intramolecular components of the vibrational spectral density can enhance coherent transport rather than destroying it. This is in contrast to the conclusions of Christensson~\textit{et al.} \cite{Christensson2012}. The vibronic theory of Refs.~\cite{Christensson2012, Chenu2012} neglects noise effects, including damping, on the vibrational mode and focuses on vibrational coherences that are localised on a single pigment. Our results show that the resonance-induced delocalisation of the mixed vibronic-excitonic states and the decay of vibrational excitations are key to the enhancement of transport. Indeed, by strongly mixing a vibrationally excited level on one site with the vibrational ground state of another site, vibration-assisted resonance creates an even greater degree of coherent delocalisation across the network than excitonic coupling alone. More significantly, the microscopic mechanism presented here may be able to conceptually reconcile the large reorganisation energies found in atomistic numerical studies~\cite{Olbrich2011c,Shim2012} with experimental observations showing the persistence of coherent oscillations up to physiological temperatures~\cite{Panitchayangkoon2010}. Further work on the full FMO complex with multiple vibrational modes and finite temperatures will address this question. 

\section{Acknowledgements}

Helpful discussions with Aggie Bra{\'n}czyk, Simon Benjamin, Erik Gauger, Kieran Higgins, Felix Pollock, Amir Fruchtman, Vaia Patta, and Elliott Levi are gratefully acknowledged. BWL thanks the Royal Society for a University Research Fellowship. This work was funded by the Leverhulme Trust. 

%\bibliography{vibres,vibres2}

\end{document}